\documentclass[12pt]{article}
\usepackage{amsmath}
\usepackage{graphicx,psfrag,epsf}
\usepackage{enumerate}
\usepackage{natbib}
\usepackage{url} 
\usepackage{algorithm}
\usepackage{algorithmicx}
\usepackage{algpseudocode}
\usepackage{tabularx}
\usepackage{booktabs}
\usepackage{amssymb}
\usepackage{amsfonts}
\usepackage{amsthm}
\usepackage{float}
\usepackage{subfigure}
\usepackage{bbm}
\usepackage{verbatim}
\usepackage{graphicx}
\usepackage{lipsum}
\usepackage{caption}
\usepackage{subcaption}
\usepackage{multirow}
\usepackage[doublespacing]{setspace}
\allowdisplaybreaks
\usepackage[margin=1in]{geometry}

\newcommand{\blind}{1}

\newtheorem{theorem}{Theorem}

\newtheorem{definition}{Definition}
\newtheorem{assumption}{Assumption}

\def\hat{\widehat}
\def\tilde{\widetilde}

\newcommand{\bo}{\boldsymbol}

\newcommand{\sT}{\mathsf{T}}

\newcommand{\D}{\mathbf D}

\newcommand{\M}{\mathbf M}

\newcommand{\R}{\mathbf R}

\newcommand{\U}{\mathbf U}
\newcommand{\V}{\mathbf V}
\newcommand{\X}{\mathbf X}
\newcommand{\Y}{\mathbf Y}
\newcommand{\Z}{\mathbf Z}

\newcommand{\bS}{\mathbf\Sigma}
\newcommand{\bbeta}{\mathbf \beta}

\newcommand{\EE}{\mathbb E}
\newcommand{\PP}{\mathbb P}

\newcommand{\hZ}{\hat{Z}}
\newcommand{\hM}{\hat{\M}}
\newcommand{\hgam}{\hat{\gamma}}

\begin{document}

\def\spacingset#1{\renewcommand{\baselinestretch}%
{#1}\small\normalsize} \spacingset{1}

\if1\blind
{
  \title{\bf Valid and Efficient Two-Stage Latent Subgroup Analysis with Observational Data}
  \author{Yuanhui Luo \\
    Department of Mathematics, Hong Kong University of Science and Technology\\
    and \\
    Xinzhou Guo\thanks{Corresponding author: Email: xinzhoug@ust.hk} \\
    Department of Mathematics, Hong Kong University of Science and Technology\\
    and \\
    Yuqi Gu\thanks{Corresponding author: Email: yuqi.gu@columbia.edu} \\
    Department of Statistics, Columbia University
    }
  \maketitle
} \fi

\if0\blind
{
  \bigskip
  \bigskip
  \bigskip
  \begin{center}
    {\LARGE\bf Valid and Efficient Two-Stage Latent Subgroup Analysis with Observational Data}
\end{center}
  \medskip
} \fi

\bigskip
\begin{abstract}
Subgroup analysis evaluates treatment effects across multiple sub-populations. When subgroups are defined by latent memberships inferred from imperfect measurements, the analysis typically involves two inter-connected models, a latent class model and a subgroup outcome model. The classical one-stage framework, which models the joint distribution of the two models, may be infeasible with observational data containing many confounders. The two-stage framework, which first estimates the latent class model and then performs subgroup analysis using estimated latent memberships, can accommodate potential confounders but may suffer from bias issues due to misclassification of latent subgroup memberships. This paper focuses on latent subgroups inferred from binary item responses and addresses when and how a valid two-stage latent subgroup analysis can be made with observational data. We investigate the maximum misclassification rate that a valid two-stage framework can tolerate. Introducing a spectral method perspective, we propose a two-stage approach to achieve the desired misclassification rate with the blessing of many item responses. Our method accommodates high-dimensional confounders, is computationally efficient and robust to noninformative items. In observational studies, our methods lead to consistent estimation and valid inference on latent subgroup effects. We demonstrate its merit through simulation studies and an application to educational assessment data.
\end{abstract}

\noindent%

\vfill

\newpage
\spacingset{1.7}

\section{Introduction}
\label{sec:intro}

Subgroup analysis, the analysis of heterogeneous treatment effects across multiple sub-populations, plays a crucial role in many scientific fields. In biomedical studies, modern therapies are often found to be helpful in some subpopulations but much less so for others. For example, \cite{amado2008wild} noted that panitumumab shows heterogeneous treatment effects in metastatic colorectal cancer patients with different subtypes of the KRAS gene. When this happens, subgroup analysis can enhance our understanding of the treatment and provide guidance on where the treatment should be implemented. Besides biomedical studies, subgroup analysis is also widely used in the broad areas of data science, such as educational studies \citep{yin2024effects} and policy learning \citep{shi2021online}. 

To conduct subgroup analysis, we first need to appropriately define subgroups, which typically fall into two categories, classical subgroups and latent subgroups. Classical subgroups refer to the subgroups defined explicitly by observed pre-treatment variables based on domain knowledge or data-driven methods \citep{loh2019subgroup}; e.g., {\it\{female\}} and {\it\{age$\geq$60\}}. In practice, classical subgroups are widely adopted due to their natural segregation of population and ease of interpretation \citep{luo2023inference}. Based on randomized trials, numerous methods have been developed to estimate and infer the treatment effects of classical subgroups \citep{guo2021inference, cai2022capital}. Although randomized trials remain the gold standard for subgroup analysis, they might not be feasible due to cost or ethical constraints \citep{hebert2002design}. In such cases, the prevalence of observational data, such as large-scale educational assessments \citep{gu2019learning}, provides unique opportunities \citep{guo2023assessing, lipkovich2023modern}. Different from classical subgroups, latent subgroups refer to the subgroups defined by unobserved latent memberships inferred from imperfect measurements; e.g., {\it\{students with strong motivation\}} inferred from students' responses to educational testing items and {\it\{depressed patients\}} inferred from mental health survey questionnaires. Latent subgroups are often adopted when directly defining classical subgroups by observed pre-treatment variables is infeasible or the subgroup of interest is an unobserved latent construct in nature \citep{lanza2013alatent}. In the presence of such latent heterogeneity, the adoption of latent subgroups can help subgroup analysis be more feasible and interpretable \citep{kim2019estimating}. 

How to make valid and efficient estimation and inference on the treatment effects of latent subgroups is a critical question in the broad areas of data science, such as educational, social, and health studies \citep{page2015principal}. One motivating example of this paper is Trends in International Mathematics and Science Study (TIMSS), a large-scale educational assessment that investigates the students’ performance in mathematics and science around the world \citep{martin2015timss}. In TIMSS, one educational question of interest is how private science lessons impact the performance of students with different motivations in science assessment \citep{lyu2023estimating}. Although student motivation is captured by about 70 binary item responses (agree or disagree) with respect to opinions on learning mathematics and sciences from TIMSS, defining $2^{70}$ classical subgroups directly based on answers to these questions is clearly neither infeasible nor interpretable. Indeed, students with different motivations are often viewed as latent subgroups inferred from imperfect measurements, the answers to questionnaire items regarding opinions on learning mathematics and sciences \citep{bechter2018student}. Moreover, TIMSS is an observational study with potentially high-dimensional confounders, such as family socioeconomic status and parent education. Such problem setups are not unique to TIMSS and also arise in other scientific disciplines, such as social science surveys and biomedical studies. For example, in studies of National Longitudinal Survey of Adolescent Health, adolescents with different risks are viewed as latent subgroups inferred from 6 binary risk characteristics \citep{lanza2013alatent}, and in studies of acute respiratory distress syndrome (ARDS) associated with COVID-19, ARDS subphenotypes are viewed as latent subgroups inferred from demographics, vital signs and laboratory values in electronic health records \citep{sinha2021latent}. Both survey and electronic health records are observational data with potentially high-dimensional confounders, such as school management and use of immunomodulatory drugs. Therefore, it is desirable to propose appropriate methods to accurately and efficiently estimate and infer the treatment effects across latent subgroups while adjusting for confounders.

Latent subgroup analysis with observational data can be much more challenging than that for classical subgroups as it typically consists of two inter-connected models \citep{zhou2022subgroup}, a latent class model and a subgroup outcome model. The latent class model relates latent subgroup memberships with multivariate imperfect measurements, such as item responses collected from assessments or surveys. The subgroup outcome model evaluates the treatment effects in latent subgroups by modeling the relationship between latent subgroup memberships, treatment, potential confounders, and outcomes of interest. To estimate and infer latent subgroup effects, some existing methods consider the classical one-stage framework, which fits two models simultaneously by maximizing the (penalized) joint likelihood \citep{shen2020subgroup, elliott2020methods, kang2023identification}. The one-stage framework is widely adopted in practice, but it is likelihood-based and might encounter several issues with high-dimensional observational data. First, the one-stage framework could be computationally infeasible or inefficient when many potential confounders or item responses are included, leading to a complex joint likelihood function and challenging optimization problems \citep{schuler2014addressing}. Second, the one-stage framework might be theoretically biased when the joint likelihood is misspecified or the parameter space is non-identifiable or increasingly complex with high-dimensional potential confounders \citep{vermunt2010latent, shen2015inference, lythgoe2019latent}. Third, the one-stage framework fits the two models simultaneously, in conflict with the common belief of practitioners that subgroup effects should be evaluated after determining subgroup memberships to bear a clear interpretation of subgroups not subject to the outcomes or confounders. Such a one-stage fitting would also increase the risk of data snooping and become infeasible when the data are stored separately \citep{vermunt2010latent}. 

Besides the one-stage framework, there is also considerable research dedicated to the two-stage framework which fits two models separately by first estimating the latent class model and then evaluating subgroup effects with estimated latent memberships \citep{bakk2021relating, kang2020causal}. The two-stage framework is intuitive and easy to implement, might accommodate many potential confounders, and has been widely adopted in the analysis of latent subgroups with observational data. Examples include the study of latent subgroup effects of preschool programs to disadvantaged children \citep{cooper2014benefits} and corticosteroids to Covid-related ARDS patients \citep{sinha2021latent}. However, as the estimated latent membership might not be the ground truth, the two-stage framework might introduce misclassification in the first stage and bias the analysis of subgroup effects in the second stage. To better support the practical use of the two-stage framework, it is desirable to understand when and how the two-stage latent subgroup analysis remains valid with observational data. 

In this paper, we focus on latent subgroups related with binary item responses through a latent class model and consider the two-stage framework for observational data with both low-dimensional and high-dimensional confounders. We first derive the maximum misclassification rates under which misclassification bias is negligible in two-stage latent subgroup analysis via a careful examination of the impacts of misclassified subjects. Then, we introduce a spectral method perspective into the two-stage framework and propose a spectral clustering-based two-stage approach to achieve the desired misclassification rate. Theoretically, the derived misclassification rate is sharp in any compact parameter space, leading to realistic design conditions for our method. Methodologically, our method is computationally efficient, robust to noninformative items, and capable of accommodating high-dimensional confounders. With the blessing of many item responses, our method gives consistent estimation and valid inference on latent subgroup effects in observational studies.

Existing two-stage approaches for latent subgroups typically fit the latent class model via the EM algorithm \citep{bakk2014relating}, which is computationally inefficient and lacks theoretical guarantees for the latent class model with many items \citep{kwon2019global}. Although several approaches have been proposed to correct the misclassification bias in the two-stage framework \citep{bolck2004estimating, bakk2018two}, they are typically designed for point estimation in the absence of covariates and are not directly applicable for inference on the subgroup treatment effects from observational data. To the best of our knowledge, rigorous studies of misclassification bias and efficient and provable two-stage approaches for latent subgroups are still lacking especially in observational studies, and we aim to bridge the gap in this paper.

In summary, this paper makes contributions in both theory and methodology. Theoretically, we derive the maximal misclassification rate and address the question of when a two-stage framework remains valid with observational data. This lays a solid foundation for proposing a valid two-stage approach. Methodologically, we integrate the spectral method perspective to latent subgroup analysis and design a valid, robust, and efficient two-stage approach with observational data in broadly applicable scenarios. 

The remainder of the paper is organized as follows. In Section \ref{sec:when}, we state the problem setting and derive the maximum misclassification rate. In Section \ref{sec:method}, we propose the spectral clustering-based two-stage approach and investigate its theoretical properties. In Section \ref{sec:simulation}, we demonstrate the merit of our method through simulation studies. In Section \ref{sec:application}, we apply our method to analyze an educational assessment dataset and study how private science lessons impact the performance of students with different motivations in the science assessment. In Section \ref{sec:discussion}, we conclude the paper with a discussion. All the proofs are presented in the Supplementary Materials.

\section{Two-Stage Latent Subgroup Analysis}\label{sec:when}
In this section, we address the question of when the two-stage framework remains valid with observational data. To start with, we introduce the latent class model and subgroup outcome model in latent subgroup analysis with observational data, and illustrate the benefits and challenges, particularly misclassification bias, of the two-stage framework. Then, we derive the maximum misclassification rate that a valid two-stage framework can tolerate in the presence of potential confounders.

\subsection{Problem Setting}

Consider an observational study of $n$ subjects with observation $\{(Y_i, D_i,  R_i, X_i)\}_{i=1}^{n}$ and unobserved latent subgroup membership $\{Z_i\}_{i=1}^{n}$. Assume the quintuplets $\{(Y_i, D_i, Z_i, R_i, X_i)\}_{i=1}^{n}$ are $n$ i.i.d. copies of $(Y, D, Z, R, X)$, where $Y$ is the outcome of interests, $D \in \{0,1\}$ is the treatment indicator, $Z \in [G]: = \{1,\ldots, G\}$ is the unobserved membership of $G$ latent subgroups, $R$ is a $J$-dimensional vector of binary item responses and $X$ is a $p$-dimensional vector of potential confounders. Under the potential outcome framework, we have $Y=(1-D)Y(0)+DY(1)$ where $Y(1)$ and $Y(0)$ denote the potential outcomes under the treatment and control, respectively. In this paper, our goal is to estimate and infer the average treatment effect of the $g$-th latent subgroup $\EE[Y(1)-Y(0)\mid Z=g]$, $g\in [G]$, based on the observed study data $\{(Y_i, D_i,  R_i, X_i)\}_{i=1}^{n}$. 

To start with, we model the outcome and relate $Y$ with $X$, $D$, and $Z$. In particular, following the subgroup analysis setup in \cite{guo2023assessing} and \cite{belloni2014inference}, we assume $Y$ can be approximated by a linear transformation of $X$ and consider a linear subgroup outcome model 
\begin{equation}\label{outcome-model}
    Y=\sum_{g=1}^G I(Z=g)\alpha_g+D \sum_{g=1}^G I(Z=g) \mu_g+X^\sT \beta+\varepsilon,
\end{equation}
where $\alpha_g$, $\mu_g$ and $\bbeta$ are the regression coefficients and $\varepsilon$ is a Gaussian error term. Under the strong ignorability assumption $\{Y(1), Y(0)\} \bot D\mid (Z,X)$, we have $\mu_g=\EE[Y(1)-Y(0)\mid Z=g]$ is the treatment effect of the $g$-th latent subgroup and $\mu=(\mu_1,\ldots,\mu_G)^\sT$ is thus the quantity of interest. When the subgroup membership $Z$ is observed, estimation and inference of the subgroup effects $\mu$ have been discussed in \cite{imai2013estimating} and \cite{guo2023assessing}. However, here, we focus on a scenario where $Z$ is latent and unobserved. 

Next, we model the latent subgroup membership and relate $Z$ with $R$. In particular, we consider a latent class model with binary item responses \citep{goodman1974exploratory,clogg1995latent} where the latent class variable $Z$ follows the categorical distribution with proportion parameters $\tau = \{\tau_g\}_{g=1}^G$; i.e., $\PP(Z=g)=\tau_g$, $g=1,\ldots, G$, and given $Z=g$, the item responses are assumed to be conditionally independent Bernoulli random variables with item parameters $\theta_g=(\theta_{1,g},\ldots,\theta_{J,g})^\sT$, $g\in [G]$; i.e.,
\begin{equation}\label{latent-model}
    \PP(R=r) = \sum_{g=1}^G \tau_g \prod_{j=1}^J \theta_{j,g}^{r_j} (1-\theta_{j,g})^{1-r_j}, \quad \forall r\in\{0,1\}^J.
\end{equation}
where $r_j$ is the $j$-th element of $r$. This modeling is different from that in fusion-type subgroup analysis \citep{ma2017concave} where subgroup membership is not related with item responses. We denote the item response matrix by $\R = \{R_i\}_{i=1}^n \in\{0,1\}^{n\times J}$ and the item parameter matrix by $\bo\Theta = \{\theta_{j,g}\}^{J\times G} \in [0,1]^{J\times G}$. 

Note that in Eqs. \eqref{outcome-model} and \eqref{latent-model}, we do not model the relationship between the potential confounders $X$ and the latent subgroup membership $Z$ or item responses $R$. Such flexibility allows the distribution of potential confounders to be heterogeneous across different subgroups and more importantly, a variable can serve the roles of both a confounder and an item response. In other words, $X$  can include item responses $R$. In addition, we consider a linear subgroup outcome model in Eq. \eqref{outcome-model} and a latent class model in Eq. \eqref{latent-model} for the sake of simplicity. Extensions to generalized linear models (beyond linear models) and other mixture models (beyond latent class model) are feasible, as long as: (a) the confounding effect can be well approximated by a (sparse) linear combination of potential confounders as discussed in \cite{wang2019debiased} and \cite{guo2023assessing}, and (b) the response data in the mixture model are sub-Gaussian as described in \cite{zhang2022leave}.

\begin{figure}[!htb]
    \centering
    \subfigure[Latent subgroup analysis]{\includegraphics[width=0.3\linewidth]{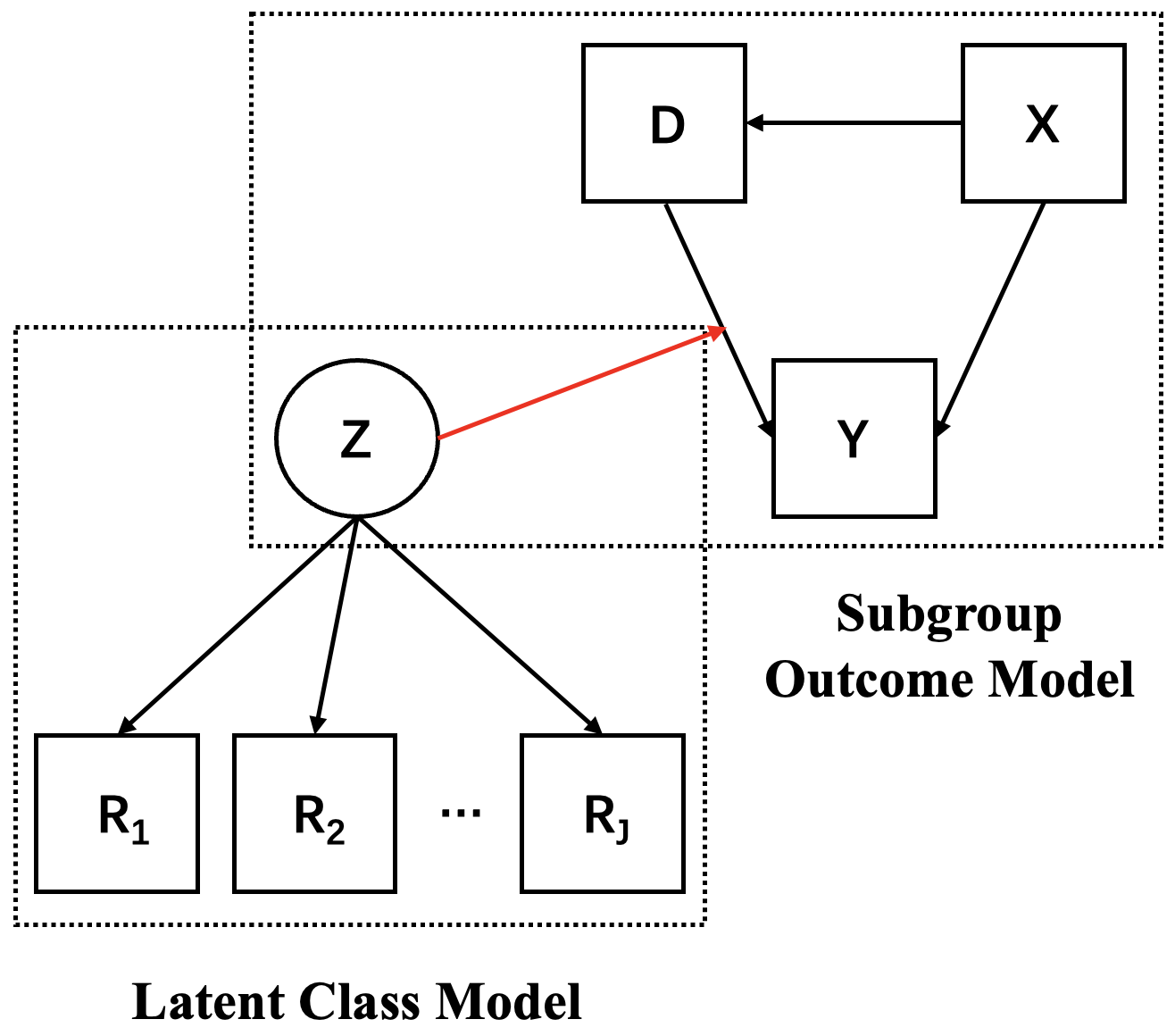}\label{fig:model}}
    \hfill
    \subfigure[One-stage framework]{\includegraphics[width=0.3\linewidth]{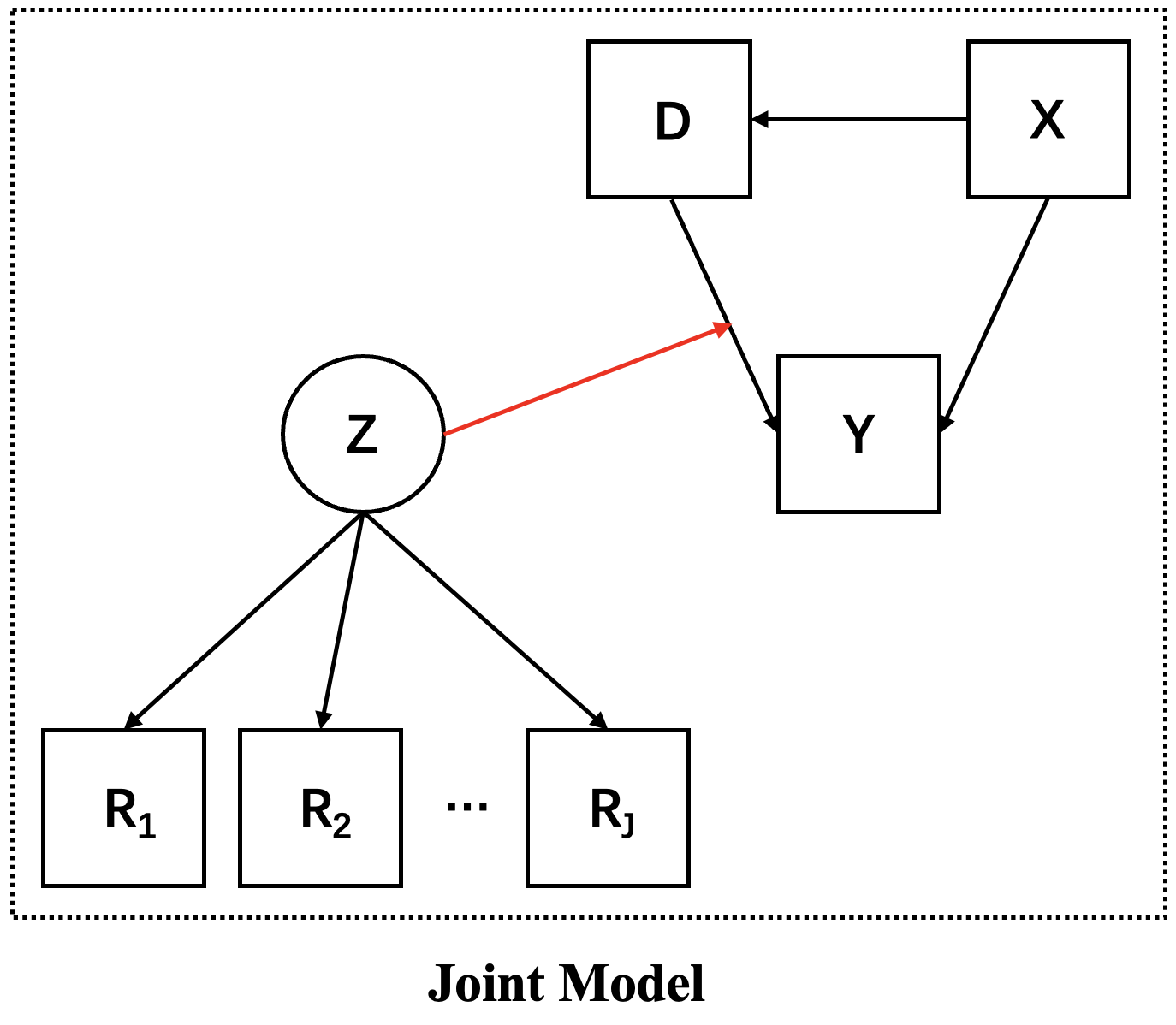}\label{fig:one_stage}}
    \hfill
    \subfigure[Two-stage framework]{\includegraphics[width=0.3\linewidth]{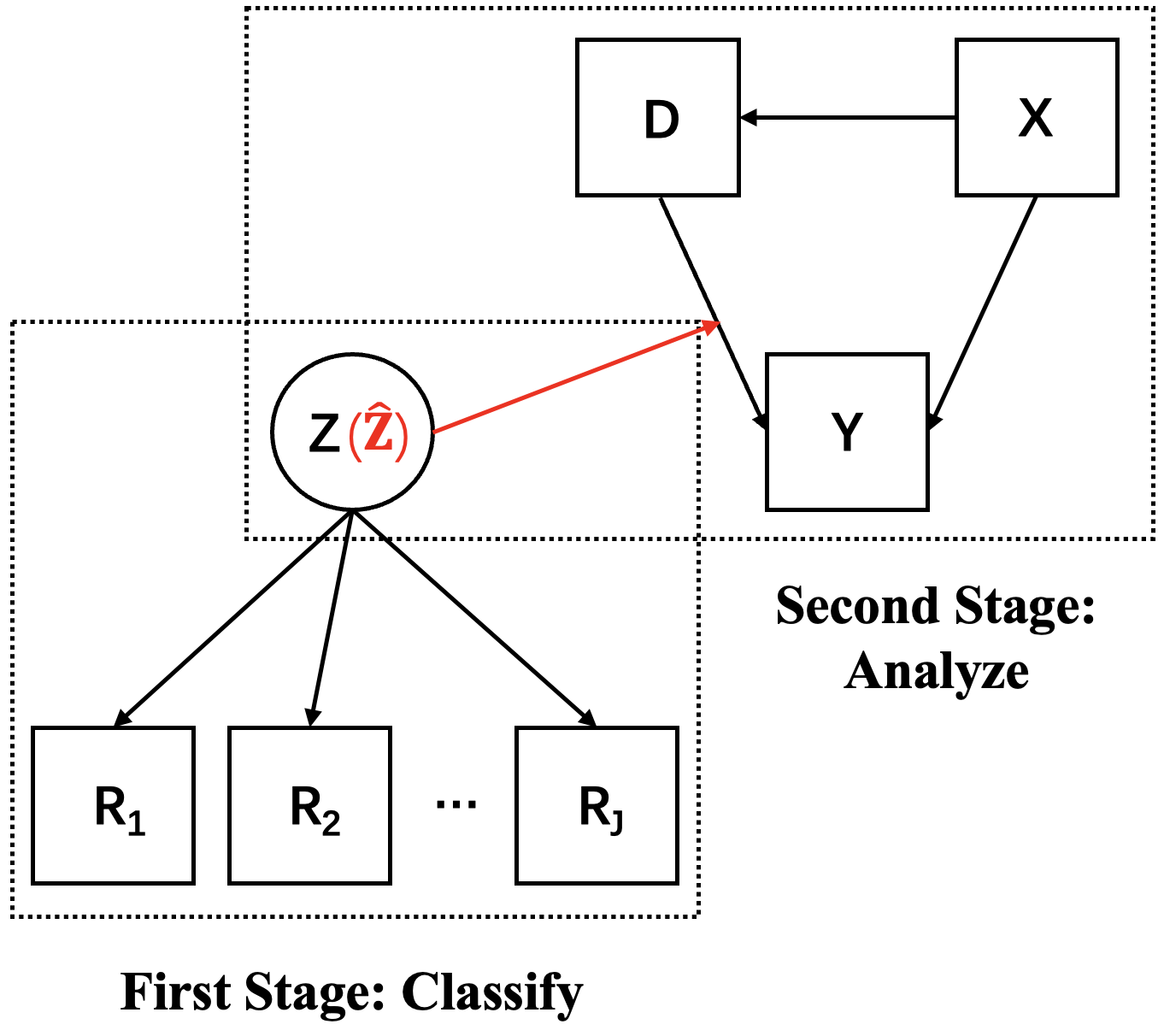}\label{fig:two_stage}}
    \caption{Illustration of two frameworks for latent subgroup analysis with observational data.}
    \label{fig:illustration}
\end{figure}

Figure \Ref{fig:illustration}\subref{fig:model} illustrates the graphical model representation of latent subgroup analysis with observational data, which consists of two inter-connected models: the latent class model and the subgroup outcome model. To estimate and infer the latent subgroup effect $\mu$, one can consider either a one-stage or a two-stage framework. The one-stage framework, as illustrated by Figure \Ref{fig:illustration}\subref{fig:one_stage}, maximizes the joint likelihood $\PP(Y,D,R,X)$ of all variables and fits the two models simultaneously. Although the validity and practical implementation of the one-stage framework have been studied for classical low-dimensional settings, it typically relies on the correct specification of $\PP(Y,D,R,X)$ and the low statistical complexity of the whole parameter space $(\tau, \bo\Theta, \alpha,\mu,\beta)$. In observational studies, such requirements are hard to satisfy as the number of potential confounders plus item responses $J+p$ is often large and the joint likelihood of the two models $\PP(Y,D,R,X)$ is often complex with numerous parameters. Moreover, simultaneously fitting two models contradicts the usual practice in subgroup analysis that subgroup effects $\mu$ are evaluated after subgroup memberships have been determined. This might even be impossible when the data for item responses $R$ and the others $(Y,D,X)$ are stored separately. To address the above issues, the two-stage framework, as illustrated by Figure \Ref{fig:illustration}\subref{fig:two_stage}, has been considered which proceeds by first estimating latent subgroup membership $\hZ$ based on the item response $R$ alone and then evaluating the latent subgroup effect $\mu$ by regressing $Y$ on $\hZ, D\hZ, X$ \citep{bakk2021relating}.

\begin{figure}[!htb]
    \centering
    \includegraphics[width=1\linewidth]{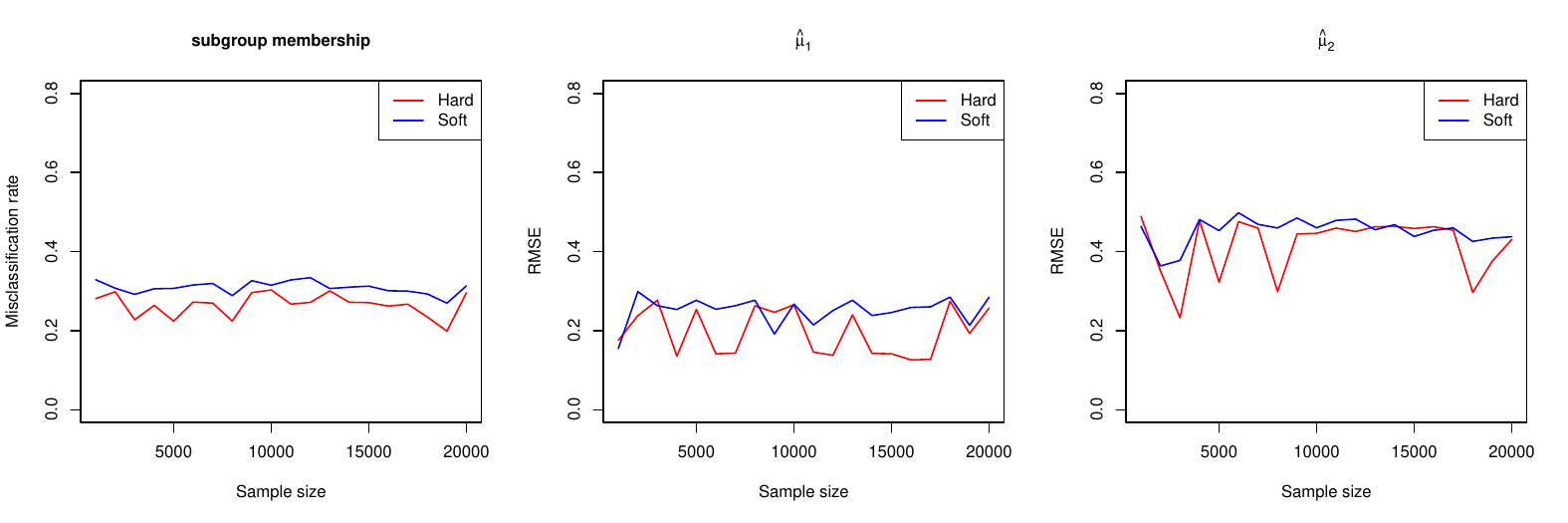}
    \caption{The misclassification rate of the estimated latent subgroup membership and the root mean square error (RMSE) of the two-stage estimators for $\mu=(\mu_1,\mu_2)^\sT$ based on hard assignment and soft assignment w.r.t. different sample sizes with $\tau=(0.6,0.4)$, $\theta_1=(0.7,0.8)$, $\theta_2=(0.3,0.2)$ and $G=2$ for latent class model in Eq. \eqref{latent-model} and $\alpha=(0,0)$, $\mu=(1,2)$, $X \sim N(0,\textbf{I}_p)$, $\beta=\textbf{1}_p$ and $\epsilon \sim N(0,0.5^2)$ under random treatment for subgroup outcome model in Eq. \eqref{outcome-model}.}
    \label{fig:toy_example}
\end{figure}

Obviating the need for jointly modeling all variables and simultaneously fitting two models, the two-stage framework provides a potentially more feasible way to conduct latent subgroup analysis with observational data. However, we must note that the estimated memberships $\hZ$ are not necessarily equal to the true $Z$ and such misclassification might introduce bias. As illustrated by the numerical example in Figure \ref{fig:toy_example}, misclassification bias might not diminish by simply considering different fitting methods of the latent class model, such as hard assignment \citep{nagin2005group} and soft assignment \citep{goodman20071}, or increasing the sample size \citep{neyman1948consistent}. This is because even when we know the true parameter of the latent class model, a subject might be misclassified with non-zero probability; i.e., $P(\hZ=g|Z=g) = \sum_{r \in \mathcal{R}} P(\hZ=g|R=r) P(R=r|Z=g) < 1$, unless $P(\hZ=g|R=r^*)=1$ for any $r^* \in \mathcal{R}$. Such intrinsic uncertainty of subgroup assignments in the first stage might substantially bias the analysis of subgroup effects in the second stage, and it is unclear whether and when the misclassification bias is negligible. 

Before moving forward, we introduce some notations for the asymptotic analysis of random variables that will be used throughout the paper. We let (1) $A_n=O_{\PP}(B_n)$ or $A_n \lesssim_{\PP} B_n$ means that there exists a constant $c>0$ such that $\lim_{n \rightarrow \infty}\PP(|A_n| \leq c|B_n|)=1$; (2) $A_n=o_{\PP}(B_n)$ means that for every constant $c>0$, $\lim_{n \rightarrow \infty} \PP(|A_n|<c|B_n|)=1$; (3) $A_n=\Omega_{\PP}(B_n)$ or $A_n \gtrsim_{\PP} B_n$ means that there exists a constant $c>0$ such that $\lim_{n \rightarrow \infty}\PP(|A_n| \geq c|B_n|)=1$; (4) $A_n=\omega_{\PP}(B_n)$ means that for every constant $c>0$, $\lim_{n \rightarrow \infty}\PP(|A_n| > c|B_n|)=1$.

\subsection{Maximum Misclassification Rate}\label{sec:misclassification}

To understand when the misclassification bias is negligible in the two-stage framework, let 
\begin{equation}\label{misclassification}
    M(\Z,\hat{\Z})=\min_{\phi\in \pi(G)} \frac{1}{n} \sum_{i=1}^n I(\hZ_i \neq \phi(Z_i))
\end{equation}
denote the misclassification rate, where the latent subgroup membership matrix $\Z = \{I(Z_i=g)\}^{n\times G}$, the estimated one  $\hat{\Z} = \{I(\hat{Z}_i=g)\}^{n\times G}$ and $\pi(G)$ refers to all permutation mappings on $[G]$. In Eq. \eqref{misclassification}, we adopt minimization over $\pi(G)$ because $\hat{\Z}$ is identifiable up to a permutation of $[G]$ \citep{loffler2021optimality}. In spectral theory literature, $M(\Z,\hat{\Z})$ is also known as misclustering error \citep{zhang2022leave}. In this paper, we adopt the name, misclassification rate, in order to make it consistent with the usual terminology used in latent subgroup analysis literature \citep{bakk2021relating}. Let $\hat{\mu}$ and $\hat{\bS}$ denote the two-stage estimator and estimated covariance matrix respectively by regressing $Y$ on $\hZ, D\hZ, X$ and $\tilde{\mu}$ denote the corresponding oracle estimator by regressing $Y$ on $Z, DZ, X$ for $\mu$ under Eq. \eqref{outcome-model}. We aim to derive the maximum misclassification rate $M(\Z,\hat{\Z})$ that a valid two-stage framework can tolerate in the presence of potential confounders. Here, validity means that $\hat{\mu}$ is consistent, leads to valid confidence intervals with $\hat{\bS}$ or achieves the oracle estimation error rate for $\mu$ as $\tilde{\mu}$, depending on the specific scenarios of interest.

Directly studying the maximum misclassification rate is challenging due to the self-correlation issue induced by $\hZ$. In particular, neither $\hZ_i$ nor $\hZ_i-Z_i$ can be treated as i.i.d. random variables, and indeed, their distributions depend on how the latent class model is fitted and the misclassification bias can be very complicated in the presence of potential confounders. Therefore, classical asymptotic theories and error-in-variable models are not directly applicable here \citep{schennach2016recent}. To derive the maximum misclassification rate, we adopt appropriate bridges between $\hat{\mu}$ and $\mu$ and quantify the impacts of misclassified subjects sharply in both low and high dimensions. 

In the low-dimensional setting where $p+2G<n$ and $p$ and $G$ are fixed, ordinary least squares (OLS),  $\hat{\mu}_{\text{ols}}$ in Eq. \eqref{eq:mu_ols} and $\hat{\bS}_{\text{ols}}$, is commonly used in fitting the subgroup outcome model for both estimation and inference on the latent subgroup effect $\mu$. In low dimensions, we adopt $\tilde{\mu}_{\text{ols}}$, the oracle OLS estimator based on $Z$, as a bridge and assume classical OLS assumption, Assumption A.1 in Appendix A, that guarantees the validity of $\tilde{\mu}_{\text{ols}}$. Then, Theorem \ref{thm:low} states that the two-stage OLS estimator $\hat{\mu}_{\text{ols}}$ is consistent and leads to valid confidence intervals with $\hat{\bS}_{\text{ols}}$ if the misclassification rate converges to 0 and goes to 0 at a rate faster than $1/\sqrt{n}$ respectively. As expected, the rate requirement for valid inference is tighter than that for consistent estimation. Moreover, Theorem \ref{thm:low} implies that any misclassification rate asymptotically larger than the desired rates will lead to inconsistent estimation or invalid inference on $\mu$ in any compact parameter space. It is in this sense that we call the desired rate the maximum misclassification rate in low dimensions.

\begin{theorem}[Maximum Misclassification Rate in Low Dimensions]\label{thm:low}
    Under Assumption A.1 required for OLS in Appendix A, we have for the two-stage OLS estimator 
    
    \noindent
    (1) (Estimation) if $M(\Z,\hat{\Z})=o_{\PP}\left(1\right)$, $\|\hat{\mu}_{\text{ols}}-\mu\|_1=o_{\PP}(1)$;
    
    \noindent
    (2) (Inference) if 
    $M(\Z,\hat{\Z})=o_{\PP}\left(\frac{1}{\sqrt{n}}\right)$, $\sqrt{n}(\hat{\mu}_{\text{ols}}-\mu)=W+\Delta_{\text{ols}}$ where $W \sim N(0,\hat{\bS}_{\text{ols}})$ and $\|\Delta_{\text{ols}}\|_{\infty}=o_{\PP}(1)$.
    
    \noindent{}Moreover, for any compact parameter space $\mathcal{S}$, there exists $(\tau_0,\bo\Theta_0,\alpha_0,\mu_0,\beta_0) \in \mathcal{S}$ under which if $M(\Z,\hat{\Z})=\Omega_{\PP}\left(1\right)$, $\|\hat{\mu}_{\text{ols}}-\mu\|_1= \Omega_{\PP}(1)$ and if $M(\Z,\hat{\Z})=\Omega_{\PP}\left(\frac{1}{\sqrt{n}}\right)$, $\|\Delta_{\text{ols}}\|_{\infty}=\Omega_{\PP}(1)$.
\end{theorem}

In the high-dimensional setting where $p+2G>n$ and $p$ and $G$ might increase with $n$, Lasso and debiased Lasso, $\hat{\mu}_{\text{lasso}}$  in Eq. \eqref{eq:mu_lasso} and $\hat{\mu}_{\text{dl}}$ in Eq. \eqref{eq:mu_dl} and $\hat{\bS}_{\text{dl}}$, are commonly used in fitting the subgroup outcome model for estimation (Lasso) and inference (debiased Lasso) on the latent subgroup effects $\mu$. In high dimensions, we can not adopt the oracle regularized estimates based on $Z$, $\tilde{\mu}_{\text{lasso}}$ and $\tilde{\mu}_{\text{dl}}$, as bridges directly, as they often lack explicit forms \citep{hastie2015statistical}. Instead, we adopt the oracle regularized loss function based on $Z$ as a bridge and bound the difference between the regularized loss function based on $Z$ and that based on $\hZ$ by the misclassification rate as detailed in Lemma 5 in Appendix A. We assume classical regularity assumptions, Assumptions A.2-A.4 in Appendix A, that guarantee the validity of $\tilde{\mu}_{\text{lasso}}$ and $\tilde{\mu}_{\text{dl}}$ in high dimensions. Then, 
Theorem \ref{thm:high} states that the two-stage Lasso $\hat{\mu}_{\text{lasso}}$ is consistent with oracle rate and the two-stage debiased Lasso $\hat{\mu}_{\text{dl}}$ leads to valid confidence intervals with $\hat{\bS}_{\text{dl}}$ if the misclassification rate converges to 0 at a rate not slower than $\sqrt{\log(p+2G)/n}$ and faster than $1/n$ respectively. Both rates are tighter than their counterparts in low dimensions to compensate for the more challenging task in high dimensions. The rate requirement for estimation is more relaxed when there are more subgroups or confounders as in that case, the oracle rate is also slower. Similar to the low-dimensional setting, the rate requirement for inference is tighter than that for estimation, and any misclassification rate asymptotically larger than the desired rates will lead to non-oracle estimation or invalid inference on $\mu$ in any compact parameter space subject to the construction of compatibility condition (Assumption A.2 in Appendix A), a classical and almost necessary condition required in high dimensions as discussed in \cite{van2017some}. It is in this sense that we call the desired rate the maximum misclassification rate in high dimensions. 

\begin{theorem}[Maximum Misclassification Rate in High Dimensions]\label{thm:high}
    Under Assumption A.2 required for Lasso and Assumptions A.2 - A.4 required for debiased Lasso in Appendix A, we have for the two-stage Lasso and debiased Lasso estimator with the usual choice of tuning parameter $\lambda \asymp \sqrt{\frac{\log(p+2G)}{n}}$
    
    \noindent
    (1) (Estimation) if $M(\Z,\hat{\Z}) = O_{\PP}\left(\sqrt{\frac{\log(p+2G)}{n}}\right)$, $\|\hat{\mu}_{\text{lasso}}-\mu\|_1=O_{\PP}\left(s_0\sqrt{\frac{\log(p+2G)}{n}}\right)$ where $s_0$ is the sparsity level of regression coefficients $\gamma=(\alpha^\sT,\mu^\sT,\beta^\sT)^\sT$;
    
    \noindent
    (2) (Inference) if 
    $M(\Z,\hat{\Z}) =o_{\PP}\left(\frac{1}{n}\right)$, $\sqrt{n}(\hat{\mu}_{\text{dl}}-\mu)=W+\Delta_{\text{dl}}$ where $W \sim N(0,\hat{\bS}_{\text{dl}})$ and $\|\Delta_{\text{dl}}\|_{\infty}=o_{\PP}(1)$.
    
    \noindent{}Moreover, subject to the construction of compatibility condition, for any compact parameter space $\mathcal{S}$, there exists $(\tau_0,\bo\Theta_0,\alpha_0,\mu_0,\beta_0) \in \mathcal{S}$ under which if $M(\Z,\hat{\Z}) = \omega\left( \sqrt{\frac{\log(p+2G)}{n}}\right)$, $\|\hat{\mu}_{\text{lasso}}-\mu\|_1 = \omega\left(s_0\sqrt{\frac{\log(p+2G)}{n}}\right)$ and if $M(\Z,\hat{\Z})= \Omega_{\PP}\left(\frac{1}{n}\right)$, $\|\Delta_{\text{dl}}\|_{\infty}=\Omega_{\PP}\left(1\right)$.
\end{theorem}

Although the maximum misclassification rates in Theorems \ref{thm:low} and \ref{thm:high} are built on OLS, Lasso and debiased Lasso respectively, extensions to other estimators, such as ridge regression, are possible as long as the estimator has an explicit form. 

\section{Valid and Efficient Two-Stage Method}\label{sec:method}
In this section, we address the question of how to design a valid two-stage framework with observational data. In specific, we introduce the spectral method perspective into the two-stage framework and propose a spectral clustering-based two-stage method to achieve the desired misclassification rate with the blessing of many item responses. We show that our method is valid, computationally efficient, and robust in broad scenarios of observational studies.

\subsection{Proposed Method}

The study of maximum misclassification rates in Section \ref{sec:misclassification} answers the question of when misclassification bias is negligible in the two-stage framework. However, it is still unclear how the desired misclassification rate in Section \ref{sec:misclassification} can be achieved in the two-stage framework. 
Although various two-stage approaches have been proposed \citep[see an overview in][]{bakk2021relating}, they often lack theoretical guarantees, especially in high dimensions, and might not be able to achieve the desired misclassification rate to infer latent subgroup effects consistently and efficiently \citep{bakk2014relating}. Specifically, existing two-stage approaches typically fit the latent class model via the Expectation-Maximization (EM; \cite{dempster1977maximum}) algorithm. Such likelihood-based procedures lack algorithmic convergence guarantees and can suffer from initialization difficulties and large computational cost, especially in the presence of a large number of items \citep{kwon2019global,balakrishnan2017statistical}.

Indeed, \cite{zeng2023tensor} showed that instead of being viewed as a challenge, the increasing availability of a relatively large number of item responses in the latent class model can help decrease the misclassification rate if the latent class model is estimated appropriately. To utilize the large number of item responses to achieve the desired misclassification rate, we introduce the spectral method perspective into the two-stage framework and particularly, adapt the spectral clustering to estimate the latent subgroup memberships in the first stage. Different from the likelihood-based method, spectral clustering groups the subjects by leveraging the approximate low-rank structure of the response data matrix and thus has the potential to efficiently accommodate high-dimensional data and capture their valuable information.
See a comprehensive monograph on spectral methods in \cite{chen2021spectral} and references therein. Spectral clustering has been extensively studied and adopted in various scenarios, such as sub-Gaussian mixture models and network models \citep{abbe2022,srivastava2023robust,zhang2022leave,li2023community}, but it has not been investigated in the latent subgroup setting considered here.
Although recent studies in \cite{lyu2024degree} and \cite{lyu2025spectral} examined spectral clustering in latent class models, they did not consider the statistically challenging downstream inference on the latent subgroup effects.
To estimate and infer the latent subgroup effect $\mu$ in a valid and efficient way, we next propose a spectral-based two-stage approach as summarized in Algorithm \ref{algo-two-stage}. For simplicity of presentation, we rewrite the subgroup outcome model Eq. \eqref{outcome-model} in matrix notation with $\Y=\{Y_i\}_{i=1}^n$, $\D=\text{Diag} \{D_i\}_{i=1}^n$ and $\X=(X_1^{\sT},\ldots,X_n^{\sT})^{\sT}$. 

\begin{algorithm}[!htb]
\caption{Two-stage Spectral Subgroup Analysis}\label{algo-two-stage}
\begin{algorithmic}[1]
\Require{Observed data $\{Y_i, D_i, R_i, X_i\}_{i=1}^n$ and the number of subgroups $G$}
\Ensure{Point estimation or confidence interval of latent subgroup effect}
\State Apply top-$G$ singular value decomposition to $\R$ and obtain $\hat{\U}_G \hat{\bS}_G \hat{\V}^{\sT}_G$, where the columns of $\hat{\U}_G$ consist of the top-$G$ left singular vectors, $\hat{\bS}_G$ is a diagonal matrix of the $G$ largest singular values of $\R$ and the rows of  $\hat{\V}_G$ consist of the top-$G$ right singular vectors;

\State Perform K-means++ based on the row vectors of $\hat\U_G \hat{\bS}_G$ and get the estimated latent subgroup memberships $\hZ_i$ for $i=1,\dots,n$:
\begin{align*}
    \left(\{\hat Z_i\}_{i\in[n]},~ \{\hat{\theta}^*_g\}_{g\in[G]}\right) = \mathop{\arg\min}_{\{ Z_i\}_{i\in[n]} \in [G], \{\theta^*_g\}_{g\in[G]} \in [0,1]^G} \sum_{i=1}^n \left\|\hat\U_{i,:}\hat{\bS}_G - \theta^*_{Z_i} \right\|^2;
\end{align*}

\State Conduct point estimation or construct a confidence interval for the latent subgroup effects by regressing $Y$ on $\hZ, D\hZ, X$, where the estimated latent subgroup effects $\hat{\mu}$ are obtained as follows 

\begin{equation}\label{eq:mu_ols}
    \hat{\mu}_{\text{ols}} \text{ by } \arg \min_{\gamma} \frac{1}{n} \left\|\Y-\hM\gamma\right\|_2^2
\end{equation}

\begin{equation}\label{eq:mu_lasso}
\hat{\mu}_{\text{lasso}} \text{ by } \arg \min_{\gamma} \frac{1}{n} \left\|\Y-\hM\gamma\right\|_2^2+\lambda \|\gamma\|_1
\end{equation}

\begin{equation}\label{eq:mu_dl}
\hat{\mu}_{\text{dl}}\text{ by }  \hgam_{\text{lasso}}+\hat{\Omega}\hM^{\sT}(\Y-\hM \hgam_{\text{lasso}})/n,  
\end{equation}

\noindent{}with $\hM:=(\hat{\Z} ~ \D\hat{\Z} ~ \X)$ and $\hat{\Z} = \{I(\hZ_i=g)\}^{n\times G}$. The definition of $\hat{\Omega}$ from nodewise Lasso and the construction of confidence intervals with $\hat{\bS}_{\text{ols}}$ and $\hat{\bS}_{\text{dl}}$ are specified in Section A.2.2 in Appendix A.

\end{algorithmic}
\end{algorithm}

In Algorithm \ref{algo-two-stage}, we first project the item responses $\R$ into a low-dimensional space of dimension $G$ via top-G SVD decomposition in step 1, and then estimate the latent group membership $\hZ$ via K-means++ based on the projected item response in step 2. In the end, we proceed to the second stage estimation or inference by regressing $Y$ on $\hZ, D\hZ, X$. The number of latent subgroups $G$ can be determined by prior knowledge or parallel analysis \citep{dobriban2019deterministic}. By leveraging the approximate low-rankness of item response matrix $\mathbf{R}$, we reduce the dimension from $J$ to $G$ before performing the K-means++ clustering. Since typically $G \ll J$, such a spectral-based approach substantially reduces the computational burden and alleviates the issue of model misspecification and local optima while preserving the useful latent information. 
This helps achieve the desired misclassification rate with the blessing of many item responses as detailed in Section \ref{sec:theory}. 

\subsection{Theoretical Justification}\label{sec:theory}

To justify the proposed two-stage spectral subgroup analysis in Algorithm \ref{algo-two-stage}, we start with an introduction of ($J_I$,$\delta$)-separability adapted from \cite{najafi2020reliable}, a key concept measuring the separation across different subgroup centers. In specific, by Definition \ref{jdelta}, ($J_I$,$\delta$)-separability requires that among all the $J$ items in the study, there are at least $J_I$ informative items whose item parameters are well separated by at least the amount of $\delta$ across different subgroups. Clearly, a larger $J_I$ or $\delta$ indicates more information contained in the item responses. 

\begin{definition}[($J_I$,$\delta$)-Separability]\label{jdelta} 
The item parameter matrix $\bo\Theta$ is said to be ($J_I$,$\delta$)-separable, if for each pair columns of $\bo\Theta$, say $g$, $g^{'}$ with $g \neq g^{'}$, there exist at least $J_I \leq J$ rows indices $\{i_1,\ldots,i_{J_I}\} \subseteq \{1,\ldots,J\}$ such that 
$$|\theta_{i_j,g}-\theta_{i_j,g^{'}}| \geq \delta, ~j=1,\ldots,J_I,$$
where $\delta \in (0,1)$.
We call the items with rows indices $\{i_1,\ldots,i_{J_I}\} \subseteq \{1,\ldots,J\}$ the $\delta$-informative items and $J_I$ the number of $\delta$-informative items. 
\end{definition}

In practice, besides informative items defined in Definition \ref{jdelta}, we might come across noninformative ones whose item parameters are not well separated across some subgroups and thus $J_I<J$. Such scenarios need to be appropriately accounted for when justifying our method. This is because, similar to other high-dimensional modeling, we usually do not know which items are informative in practice and tend to include as many items as possible in the study design. 
Although the inclusion of informative items provides a stronger signal and helps reduce misclassification bias, the inclusion of non-informative ones might introduce noise and increase misclassification bias under certain scenarios. To characterize the desired signal-to-noise ratio, we make the following assumption. Assumption \ref{apt:order} requires that (1) the sample size is asymptotically not smaller than the square of subgroup number and (2) the number of informative items is asymptotically lower bounded when the total number of items is not asymptotically larger than the sample size or the proportion of informative items is asymptotically lower bounded when the number of items is asymptotically larger than the sample size. Both requirements are easily satisfied in practice. In particular, when the number of latent subgroups $G$ and separability $\delta$ are fixed, we only need (1) the number of sample size goes to infinite and (2) the number of informative items to be a constant if $J = O(n)$ or its proportion be asymptotically not smaller than $1/n$ if  $J=\omega(n)$.

\begin{assumption}[Signal-to-Noise Ratio]\label{apt:order}
    For the $(J_I,\delta)$-separable item parameter matrix $\Theta$ with $J$ items and $G$ subgroups, we have

    \noindent
    (1) $n \gtrsim G^2$;

    \noindent
    (2) $J_I \gtrsim \frac{G^2}{\delta^2}$ when $J = O(n)$ or $\frac{J_I}{J} \gtrsim \frac{G^2}{\delta^2 n}$ when $J = \omega(n)$.
\end{assumption}

\begin{assumption}[Balanced Subgroups]\label{apt:balance}
There exists an absolute constant $C_s \in (0,1)$ such that $n_{\min}/n_{\max} \geq C_s$, where $n_{\min}=\min_{g \in [G]}|\{i: Z_i=g\}|$ and $n_{\max}=\max_{g \in [G]}|\{i: Z_i=g\}|$.
\end{assumption}

\begin{assumption}[Balanced Singular Values]\label{apt:w}
$w:=\|\bo\Theta\|_2/\sigma_{\min}(\bo\Theta)=O(1)$ with high probability, where $\sigma_{\min}(\bo\Theta)$ is the smallest non-zero singular value of $\bo\Theta$.
\end{assumption}

To justify our method, besides the signal-to-noise ratio, Assumptions \ref{apt:balance} and \ref{apt:w} introduce two mild requirements on balanced subgroups and singular values, respectively. Assumption \ref{apt:balance} states that the sizes of latent subgroups should be balanced and is easily satisfied in practice, such as congenital depressed and acquired depressed patient subgroups studied in \cite{loffler2021optimality}. Assumption \ref{apt:w} states that the singular values of the item parameter matrix have a similar magnitude, which is a reasonable assumption under common data generative schemes in latent class models following \cite{lyu2024degree}. For example, Assumption \ref{apt:w} is satisfied with high probability when $\theta_{j,g} \overset{i.i.d.}{\sim} U[0,1]$ for all $j \in [J]$ and $g \in [G]$. Under Assumptions \ref{apt:order}-\ref{apt:w}, we show how the misclassification rate decays with respect to the number of informative items $J_I$ in Theorem \ref{thm:rate}.

\begin{theorem}[Upper Bound of Misclassification Rate]\label{thm:rate}
    For the latent class model with ($J_I$,$\delta$)-separable item parameter matrix $\bo\Theta$, under Assumptions \ref{apt:order}-\ref{apt:w}, we have 
    $\mathbb{E}[M(\Z,\hat{\Z})] \leq \kappa$
    where $\kappa=O\left(\exp(-J_I \delta^2) + \exp(-n)\right)$.
\end{theorem}

Theorem \ref{thm:rate} shows that the expected misclassification rate is upper bounded by $\kappa$ and $\kappa$ decays exponentially with the number of $\delta$-informative items $J_I$ in the study design. Such an exponential decay rate allows us to justify the proposed two-stage spectral subgroup analysis across broad practical scenarios in both low dimensions and high dimensions in Table \ref{tab:conditions}, respectively. In particular, in low dimension, Table \ref{tab:conditions} states that adopting our method, the two-stage OLS $\hat{\mu}_{\text{ols}}$ is consistent when the exponential information $\exp(J_I\delta^2)\to\infty$, and leads to valid confidence intervals when  $\exp(J_I\delta^2)\to\infty$ at a rate faster than $\sqrt{n}$. When the separability $\delta$ is a constant, we only need the number of informative items $J_I\to\infty$ for low-dimensional estimation and $J_I\to\infty$ at a rate faster than $\log(n)$ for low-dimensional inference. In high dimension, Table \ref{tab:conditions} states that adopting our method, the two-stage Lasso $\hat{\mu}_{\text{lasso}}$ is consistent with oracle rate when the exponential information $\exp(J_I\delta^2)\to\infty$ at a rate not slower than $\sqrt{\frac{n}{\log(p+2G)}}$ and the two-stage debiased Lasso $\hat{\mu}_{\text{dl}}$ leads to valid confidence intervals when  $\exp(J_I\delta^2)\to\infty$ at a rate faster than $n$ respectively. When the separability $\delta$ is a constant, we only need the number of informative items $J_I\to\infty$ at a rate of $\log(n/log(p+2G))$ for high-dimensional estimation and faster than $\log(n)$ for high-dimensional inference, respectively. 

From Table \ref{tab:conditions}, we note that as expected, the design conditions are more stringent in high dimensions than low dimensions, and more stringent to achieve valid inference compared to consistent estimation. All design conditions are easily satisfied in practice, as even the most stringent one, that for high-dimensional inference, only requires the number of informative items $J_I$ to increase at a rate faster than $\log(n)$ when the separability $\delta$ is a constant. Moreover, because the desired misclassification rates in Theorems \ref{thm:low} and \ref{thm:high} are sharp and the exponential rate in Theorem \ref{thm:rate} is optimal for the mixture of isotropic Gaussians under signal-to-noise conditions as discussed in \cite{zhang2022leave}, to this extent, the desired study design conditions in Table \ref{tab:conditions} are nearly optimal. 

\begin{table}[htbp]
\centering
\caption{Latent class model design conditions. \label{tab:conditions}}
\begin{tabularx}{\textwidth}{l l X X}
\toprule
Regime & Method & Estimation & Inference \\
\midrule
\multirow{1}{*}{Low-dim (Thm 1)} 
& OLS 
& $\hat{\mu}_{\mathrm{ols}}$ is consistent if $\exp(J_I\delta^2) = \omega(1)$ 
& $\hat{\mu}_{\mathrm{ols}}$ is asymptotically normal if $\exp(J_I\delta^2)=\omega(\sqrt{n})$ \\
\multirow{1}{*}{High-dim (Thm 2)} 
& Lasso / Debiased Lasso 
& $\hat{\mu}_{\mathrm{lasso}}$ achieves oracle rate if $\exp(J_I\delta^2) = \Omega\left(\sqrt{\dfrac{n}{\log(p+2G)}}\right)$
& $\hat{\mu}_{\mathrm{dl}}$ is asymptotically normal if 
$\exp(J_I\delta^2)=\omega(n)$ \\
\bottomrule
\end{tabularx}

\vspace{0.25em}
\footnotesize\emph{Notes:} Results hold under Assumptions 1–3 and the corresponding assumptions of Theorems 1 and 2. 
\end{table}

\section{Simulation Studies}\label{sec:simulation}

In this section, we conduct simulation studies to assess the finite sample performance of the proposed two-stage spectral latent subgroup analysis. We start with a scenario where all items are informative, and consider $G=3$ latent subgroups with the total sample size $n=150$. In the latent class model, we use a similar setup as \cite{lyu2025spectral}
and consider balanced latent class proportions; i.e., $\PP(Z=g)=1/G$. We generate the (informative) item responses $\R$ by Eq. \eqref{latent-model} with the well separated item parameter matrix $\bo\Theta$, which vertically stacks $J/5$ copies of the following $5\times 3$ base matrix $\bo\Theta_{b}$:
$$
\bo\Theta_{b}
=
\begin{pmatrix}
0.25 & 0.75 & 0.75\\
0.25 & 0.75 & 0.25\\
0.75 & 0.25 & 0.75\\
0.75 & 0.25 & 0.25\\
0.75 & 0.75 & 0.25
\end{pmatrix}.
$$
We vary the number of (informative) items across $J_I=J=5,10,20,\ldots,150$. In the subgroup outcome model, we follow the data generation in \cite{shinkyu2025small}, and consider the $p$-dimensional potential confounders $X_i \sim N(0, \bS)$ with $p=10$ and $\bS_{j,k}=0.1^{|j-k|}$ for the low-dimensional setting and $p=200$ for the high-dimensional setting. The treatment indicator $D$ follows logistic distribution $\mathbb P(D=1|X)=1/(1+\exp\{-(1+X^{\sT}\beta_l)\})$ with coefficient $\beta_l=(1,1,1,1,1,1,0,\ldots,0)$, and the outcome of interest $Y$ is generated by Eq. \eqref{outcome-model} with the latent subgroup treatment effect $\mu=(-1,0,1.5)$, main effect $\alpha=(0,-0.05,-0.2)$, sparse regression coefficient $\beta=(1,1.5,1.5,0,\ldots,0)^{\sT}$ and random error $\varepsilon\sim N(0,1)$.

\begin{figure}[h!]
    \centering
    \includegraphics[width=1\linewidth]{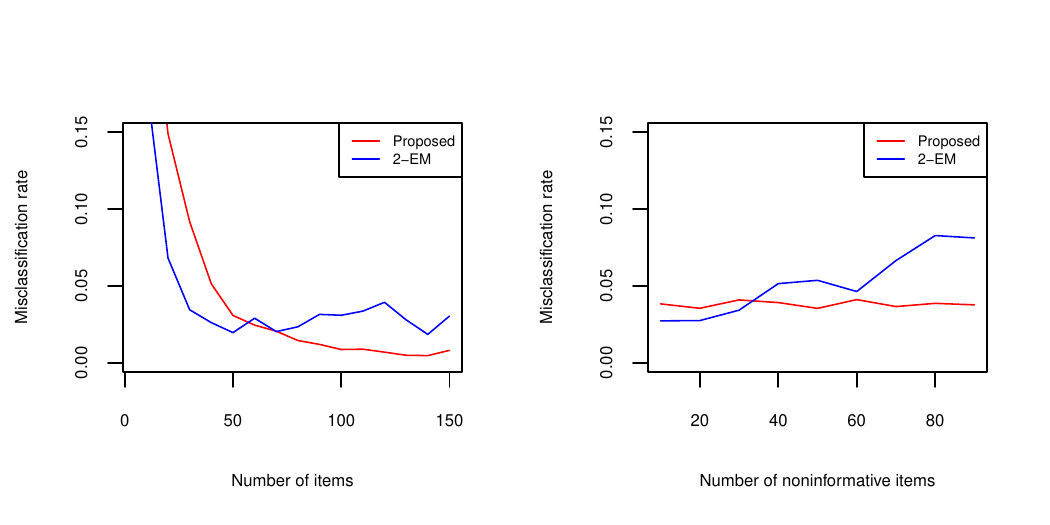}
    \caption{The misclassification rate of estimated latent subgroup membership w.r.t. different numbers of informative items (left panel) and noninformative items (right panel).}
    \label{fig:misclassification}
\end{figure}

\begin{figure}[h!]
    \centering
    \includegraphics[width=1\linewidth]{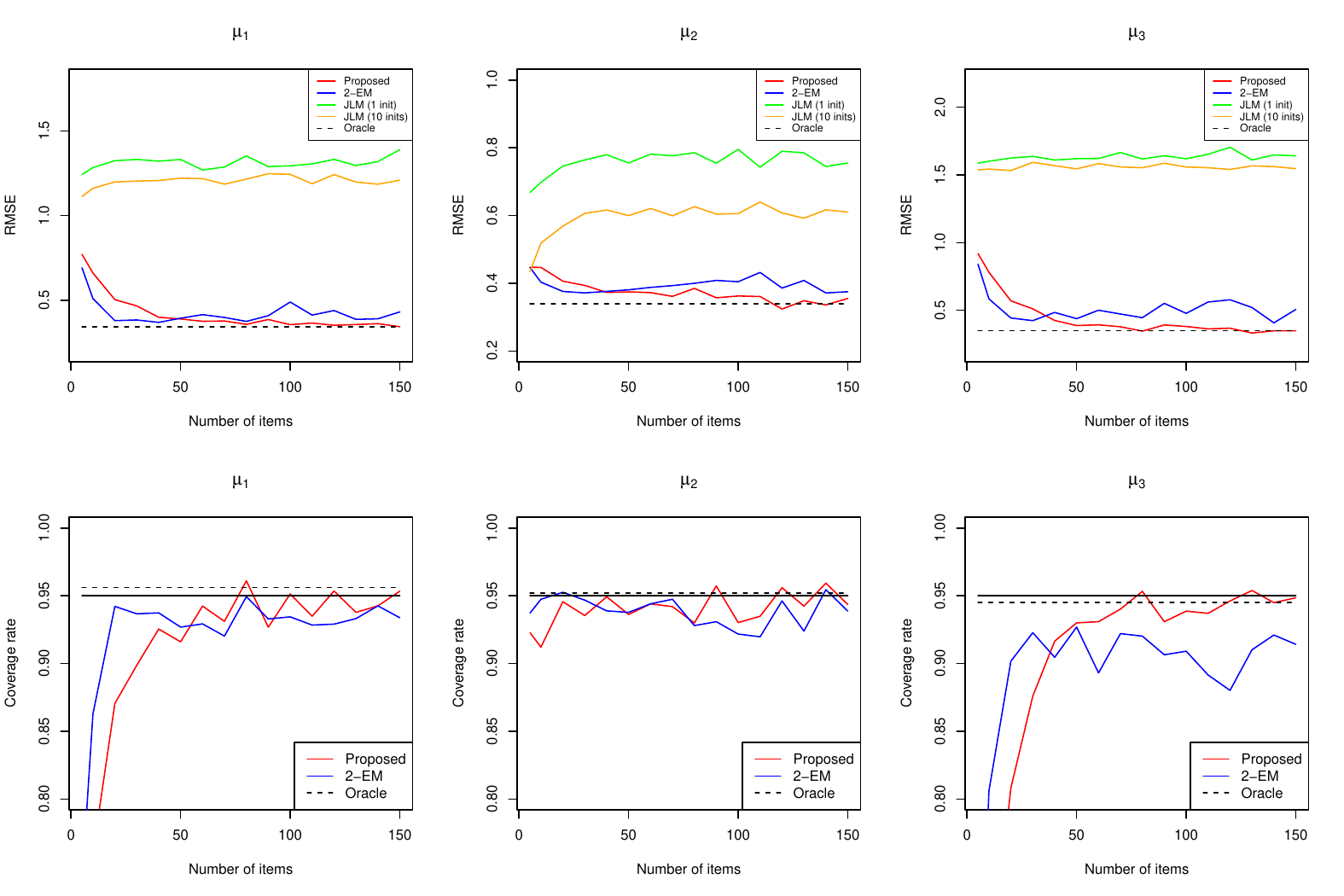}
    \caption{The root mean square estimation error and the empirical coverage probability of 95\% confidence interval with standard error below 0.01 on the treatment effect of latent subgroups w.r.t. different numbers of (informative) items when OLS is adopted in low dimension. The average computational times (seconds) of our method, two-step-EM, and joint likelihood modeling with one initialization and ten initializations are 0.076, 2.284, 17.618, and 175.246, respectively, which were conducted on a MacBook with a single M2 CPU and 24GB memory.}
    \label{fig:low}
\end{figure}

\begin{figure}[h!]
    \centering
    \includegraphics[width=1\linewidth]{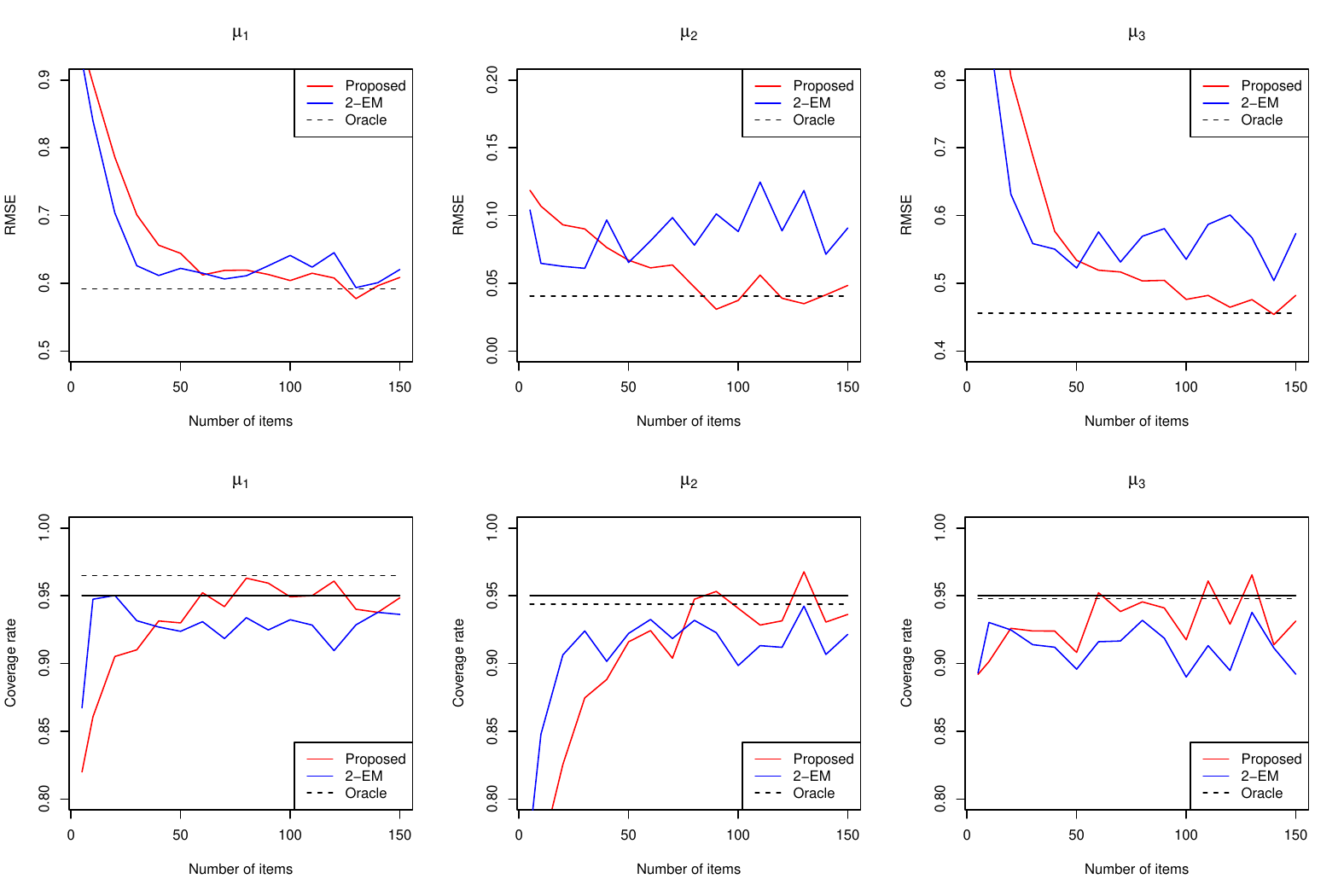}
    \caption{The root mean square estimation error when Lasso is adopted and the empirical coverage probability of 95\% confidence interval with standard error below 0.01 on the treatment effect of latent subgroups when debiased Lasso is adopted w.r.t. different numbers of (informative) items in high dimension. The average computation times (seconds) of our method, and two-step-EM are 0.314 and 2.432 for Lasso and 1.158 and 3.229 for debiased Lasso, respectively, which were conducted on a MacBook with a single M2 CPU and 24GB memory.}
    \label{fig:high}
\end{figure}

We implement the proposed two-stage spectral latent subgroup analysis with the algorithm as detailed in Section \ref{sec:method}. In particular, we assume $G$ is known, and adopt ordinary least squares for both estimation and inference in low dimensions. We adopt Lasso and debiased Lasso with cross-validated tuning parameters for estimation and inference, respectively, in high dimensions \citep{shinkyu2025small}. 
For comparison, we also implement the following methods:
\begin{enumerate}
    \item Oracle: fitting the subgroup outcome model in Eq. \eqref{outcome-model} with the true latent subgroup membership;
    \item Two-step-EM (2-EM): the two-stage approach estimating the latent subgroup membership by first running the EM algorithm with R package \texttt{poLCA} and then assigning hard labels in the first stage;
    \item Joint Likelihood Modeling (JLM): the one-stage approach fitting the joint likelihood of the two models with the EM algorithm by 1 random initialization or 10 random initializations. 
\end{enumerate}
We only consider the one-stage JLM in low-dimensional estimation as the one-stage framework is not directly applicable to high-dimensional data. One immediate drawback of JLM is that it can not well accommodate high-dimensional parameters because the complex joint likelihood challenges the parameter estimation in terms of both accuracy and efficiency. Based on 1000 independent replicates in each simulation setting, we report the misclassification rate, root mean square error for estimating the subgroup effects, empirical coverage of 95\% confidence interval for the subgroup effects, and the average computation time in Figures \ref{fig:misclassification}, \ref{fig:low} and \ref{fig:high}. 

The left panel of Figure \ref{fig:misclassification} shows that as the number of items increases, the misclassification rate of spectral clustering converges to zero as suggested in Section \ref{sec:theory}, while that of the two-stage EM method decays at a clearly slower rate and does not even converge to 0. This is because the increasing number of items complicates the parameter space and makes the EM algorithm struggle.
From Figures \ref{fig:low} and  \ref{fig:high}, we see that the proposed spectral-based method clearly outperforms the two-stage EM and one-stage EM with respect to both estimation and inference of the latent subgroup effects in both low dimensions and high dimensions. In particular, the two-stage EM and one-stage EM methods can not lead to consistent estimation and valid inference regardless of the number of items due to the limitations of the likelihood-based method. In contrast, the proposed spectral-based method indeed delivers valid estimation and inference, and its root-mean square error and empirical coverage are comparable to those of the oracle method, when there are more than 50 items in low dimensions and 100 items in high dimensions. Both are considered realistic numbers of items in practice. Moreover, our method is computationally much more efficient than the alternative methods as demonstrated by the computational time. The average computational times of JLM with one initialization and ten initializations are 17.618 and 175.246 seconds, while those of our method and two-stage EM are 0.076 and 2.284 seconds.

Next, we consider the scenario where noninformative items arise. Specifically, we consider the same data generation mechanism as the informative one with $J_I=50$ but add varying numbers of noninformative items ranging from 0 to 100 with the same item parameter for each latent subgroup; i.e. $\theta_{j,g}=0.25$ for $g \in [G]$. Figure \ref{fig:misclassification} shows that the misclassification rate of the two-stage EM clearly increases as the number of noninformative items increases, because the inclusion of noninformative items increases the complexity of parameter space and decreases the signal-to-noise ratio. On the contrary, the misclassification rate from our method remains robust with the increasing number of noninformative items as suggested in Section \ref{sec:theory}. The robustness of our method with respect to noninformative items is important in conducting latent subgroup analysis with observational data, where many items are often available but we do not know which ones are informative. 

We also conduct more simulation studies to demonstrate that (1) the one-stage approach has poor performance even with good initialization; (2) the performance of our method is robust to different sparsity levels of the coefficients $\beta$. See Appendix B for details about the additional simulation studies.

\section{Application to TIMSS Educational Assessment Data}\label{sec:application}

In this section, we apply our method to analyze a dataset from the Trends in International Mathematics and Science Study (TIMSS). TIMSS is an observational study investigating the performance of students in mathematics and science at the fourth and eighth grades in more than 50 participating countries and areas around the world since 1995. Besides students' achievement scores, the TIMSS has collected via questionnaires a rich array of students' demographic information and students' agreement levels to statements of motivation for learning mathematics and science, and educational programs students attended. Over the past decades, TIMSS has been widely used to monitor curricular implementation and identify promising instructional practices. In particular,  \cite{lyu2023estimating} identified that private science lessons might have heterogeneous effects on achievement scores of science in the latent subgroups of students with different motivations. However, such analyses using Bayesian joint estimation are based on the one-stage framework of latent subgroup analysis with observational data, which might be inadequate to adjust for the high-dimensional potential confounders.

Here, we focus on 2271 eighth-grade students of TIMSS from Hong Kong in 2015 and aim to uncover among these students, how private science lessons impact the science assessment performance of students with different motivations in science and mathematics. We let the treatment indicator $D$ denote whether or not a student has attended private science lessons during the last 12 months. Let the outcome of interest $Y$ denote the scaled scores of students' science achievement. The latent subgroups of interest $Z$ are student subgroups with different motivations. The item responses $R$ are the binary answers (agree or disagree) to $J=73$ statements about motivations for learning mathematics and science in the collected questionnaires. The potential confounders $X$ consist of the demographic information and the binary answers to statements, and their interaction terms. For the answers to statements in potential confounders, we follow the practice in \cite{lyu2023estimating} and assign scores of $0$, $\frac{1}{3}$, $\frac{2}{3}$ and $1$ to the answers "Disagree a lot", "Disagree a little", "Agree a little" and "Agree a lot", respectively to quantify the students' attitude to science and mathematics. 
The total number of potential confounders in this study is high-dimensional with $p=6676$. Previous studies of the TIMSS are typically based on the one-stage framework and thus can not accommodate many confounders. For example, \cite{lyu2023estimating} only includes  18 covariates in their study. As discussed in \cite{pearl2022comment}, a limited number of confounders might hinder the interpretation of study conclusions. Here, in the presence of high-dimensional confounders, we implement the proposed two-stage spectral latent subgroup analysis in the same way as that for high dimensions in Section \ref{sec:simulation} and choose the number of latent subgroups by parallel analysis \citep{dobriban2019deterministic}.

\begin{figure}[h!]
    \centering
    \includegraphics[width=0.9\linewidth]{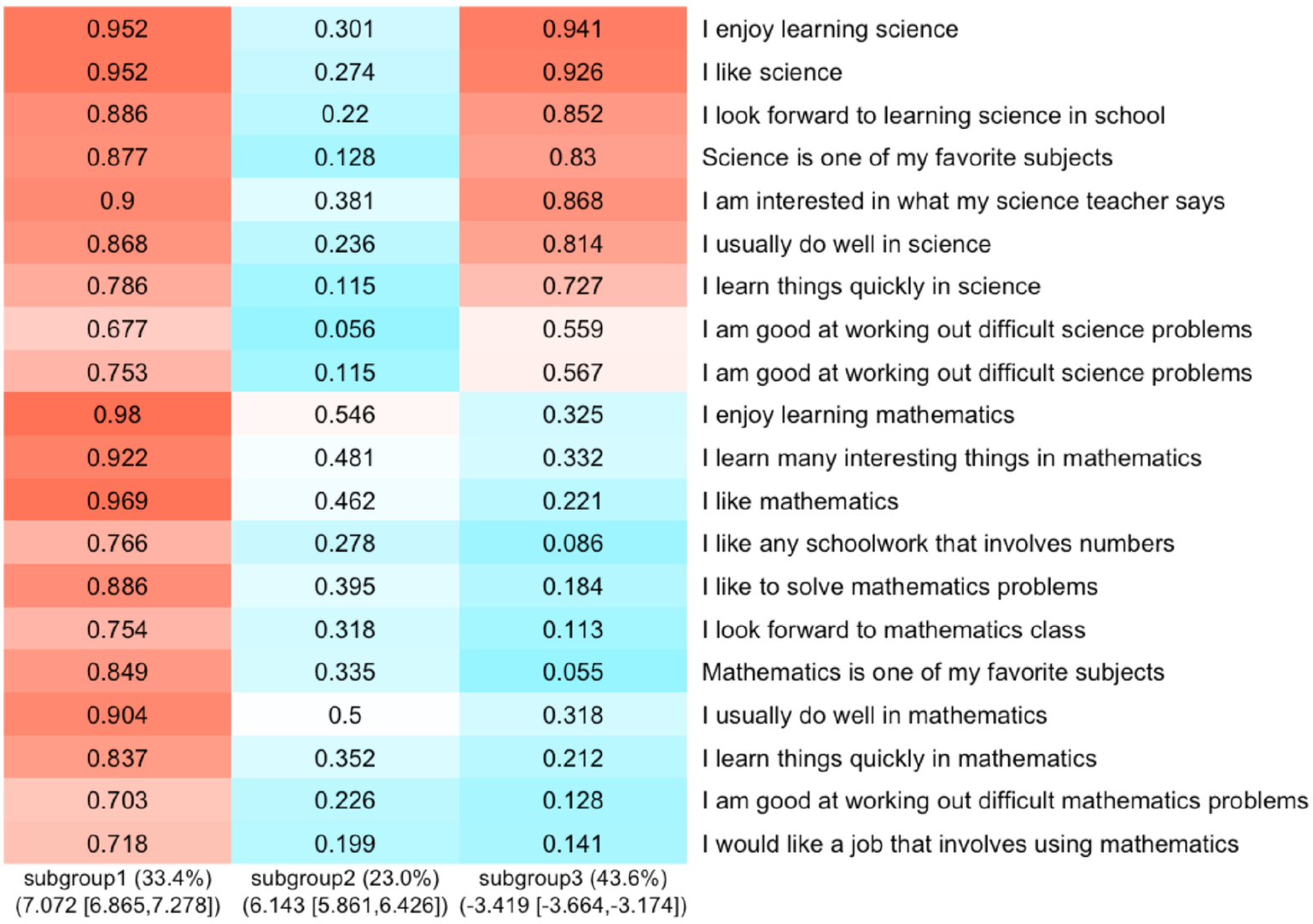}
    \caption{Heatmap of the estimated item parameter matrix $\hat{\bo\Theta}$ of 20 items with the largest variability. The darker red indicates the larger probability of agreeing with the statement while the darker blue indicates the opposite. Three latent subgroups are found with $\hat{\tau}=(0.334, 0.23,0.436)$ and $\hat{\mu}=(7.072, 6.143, -3.419)$ with 95\% confidence intervals $([6.865, 7.278], [5.861, 6.426], [-3.664, -3.174])$.} 
    \label{fig:application}
\end{figure}

Figure \ref{fig:application} 
presents a heatmap of the estimated item parameter matrix for the top 20 items with the largest variance across different subgroups. Besides the estimated item parameters, the point estimates and 95\% confidence intervals for latent subgroup effects are also presented. Based on the heatmap, we identify three latent subgroups with the following interpretations: students motivated for both science and mathematics (subgroup 1), students motivated for neither science nor mathematics (subgroup 2), and students motivated for science but not mathematics (subgroup 3). Based on the estimates and 95\% confidence intervals, attending private science lessons has positive impacts in enhancing the science achievements of students in subgroup 1 with estimated subgroup effect $7.072$ and 95\% confidence intervals $[6.865, 7.278]$ and subgroup 2 with estimated subgroup effect $6.143$ and 95\% confidence intervals $[5.861, 6.426]$, but shows negative impacts in subgroup 3 with estimated subgroup effect $-3.419$ and 95\% confidence intervals $[-3.664, -3.174]$. Possible explanations are as follows. First, for highly-motivated students in subgroup 1, private science lessons provide opportunities to further improve science achievement. Second, for unmotivated students in subgroup 2, private science lessons help identify and remediate their deficiencies in science. Third, for students adequately motivated for science but lacking interest in mathematics in subgroup 3, private science lessons nevertheless waste time of the students because their main barrier to improving science achievement is their lack of interest in mathematics. Based on this TIMSS data analysis, to improve science achievement, we would suggest the eighth grade students in Hong Kong motivated for science but not mathematics spend more time reviewing their mathematics courses instead of attending private science lessons, while encouraging the other students to take private science lessons.

\section{Discussion}\label{sec:discussion}

In this paper, we have rigorously studied the maximum misclassification rate for a valid two-stage framework and proposed a two-stage spectral latent subgroup analysis method to estimate and infer latent subgroup effects consistently and efficiently with observational data. Our method can accommodate high-dimensional potential confounders and a large number of items. We apply our method to an educational assessment dataset and reveal the latent heterogeneous effects of private science lessons on the scientific performance of students with different motivations.

In this paper, we consider a widely seen two-stage scenario where latent subgroup memberships are treated as covariates in the subgroup outcome model. In practice, latent subgroup memberships might be treated as the outcome of interest in the subgroup outcome model, particularly in many classification problems. For example, political scientists are interested in using demographic covariates to predict subgroups of ideological orientations inferred from survey responses \citep{wright2023beyond}, and market analysts are interested in using business records to predict subtypes of consumers inferred from purchase patterns \citep{wedel2000market}. When latent subgroup membership is treated as the outcome of interest, it is important to understand when and how the two-stage framework remains valid. We leave this interesting direction for future studies.

\spacingset{1}
\bibliographystyle{apalike}
\bibliography{ref.bib}

\end{document}